\documentclass[12pt]{article}
\usepackage{amsmath} 
\usepackage{epsfig} 
\input{epsf} 
\def\b0{b_0}

\def\beq{\begin{equation}} 
\def\eeq{\end{equation}} 
\def\beeq{\begin{eqnarray}} 
\def\eeeq{\end{eqnarray}} 
 
\def\to{\rightarrow}
 
\def\nn{\nonumber}

\def\gev2{{\rm GeV}^2}
\def\nn{\nonumber}

\def\lapproxeq{{\ \lower 0.6ex \hbox{$\buildrel<\over\sim$}\ }}
\def\gapproxeq{{\ \lower 0.6ex \hbox{$\buildrel>\over\sim$}\ }}

\def \be  {\begin{equation}}
\def \ee  {\end{equation}}
\def \ba  {\begin{eqnarray}}
\def \ea  {\end{eqnarray}}
\def \baa {\begin{eqnarray*}}
\def \eaa {\end{eqnarray*}}
\def \bb  {\begin{thebibliography}}
\def \eb  {\end{thebibliography}}

\begin{document} 

\begin{titlepage}
\renewcommand{\thefootnote}{\fnsymbol{footnote}}
\begin{flushright}
     \end{flushright}
\par \vspace{10mm}

\begin{center}
{\Large \bf
Seven parton amplitudes from recursion relations}
\end{center}
\par \vspace{2mm}
\begin{center}
{\bf Daniel de Florian\footnote{email address: deflo@df.uba.ar} and Jos\'e Zurita\footnote{email address: jzurita@df.uba.ar}}\\

\vspace{5mm}

Departamento de F\'\i sica, FCEYN, Universidad de Buenos Aires,\\
(1428) Pabell\'on 1 Ciudad Universitaria, Capital Federal, Argentina\\
\vspace{2mm}

\vspace{5mm}

\end{center}

\par \vspace{2mm}
\begin{center}

 {\large \bf Abstract}\\
\vspace{1cm}

We present the first calculation of  two-quark and five-gluon tree amplitudes using on-shell 
recursion relations. These amplitudes are needed for tree level 5-jet cross-section and an essential ingredient for next-to-leading order 4-jet and next-to-next-to-leading order 3-jet 
production at hadronic colliders. 
Very compact expressions for all possible helicity configurations are provided, 
allowing for direct 
implementation in Monte-Carlo codes.

 \end{center}
\begin{quote}
\pretolerance 10000

\end{quote}

\vspace*{\fill}
\begin{flushleft}
April 2006

\end{flushleft}
\end{titlepage}

\setcounter{footnote}{1}
\renewcommand{\thefootnote}{\fnsymbol{footnote}}

\section{Introduction}
\label{sec:intro}
 With the high rate for Standard Model QCD background processes at the forthcoming Large 
 Hadronic Collider (LHC), the calculation of multi-jet(particle) production cross-sections becomes 
an essential tool for the discovery of new physics.
In order to achieve a good precision within the framework of perturbative QCD, 
next-to-leading order (NLO) calculations are usually needed.
 This is in general a very complicated task in the case of hadronic colliders due to the 
increasing number of involved partons and, for that reason, such a level of accuracy has been 
reached only for a few processes.
 In several the cases, one still has to rely on tree level calculations.
 It turns out that even at the lowest order the situation starts to be complicated at the level 
of 5-jets, involving the calculation of a few thousand Feynman diagrams.

After profiting from the great simplifications \cite{Mangano:1990by,Dixon:1996wi} coming out of 
the combination of the helicity method \cite{Xu:1986xb,Kleiss:1985yh,Gunion:1985vc}
and the application of color decomposition rules for amplitudes 
\cite{Berends:1987cv,Mangano:1987xk}, the task can be performed using automatized algorithms, 
like MadEvent \cite{Maltoni:2002qb}. Even though feasible, that implies 
some non-negligible CPU time for the computation of a few million events, as can be needed for 
simulations. Counting with analytical compact expressions for the amplitudes would certainly be 
a solution for this serious inconvenience.

Fortunately, the situation has drastically improved during the last couple of 
years. After the pioneering proposal of Witten \cite{Witten:2003nn}  about the relation between 
tree level amplitudes and strings in twistor space, it was possible to formulate a set of rules 
to compute gauge amplitudes by simple recursion relations involving only 
``scalar propagators'' and the maximally helicity-violating (MHV) amplitudes, those where only 
two particles have a different polarization from the rest. Strictly speaking the first set of 
relations presented by Cachazo, Svr\v{c}ek and Witten (CSW) \cite{Cachazo:2004kj} involved the 
off-shell continuation of the MHV amplitudes, situation improved by the proposal of
Britto, Cachazo and Feng (BCF) \cite{Britto:2005fq} (later confirmed by the same authors and 
Witten \cite{Britto:2004ap}), after showing that the usual on-shell MHV amplitudes become the 
key ingredient when complex continuation of some of the external momenta is allowed.
The simplicity of the BCFW method allows to obtain very compact expressions for those amplitudes, 
explicitly exposing the high degree of symmetry hidden in the framework of direct Feynman diagram 
calculations.

The new method, initially considered for pure gluon amplitudes, has been successfully extended to 
account for the presence of massless quarks \cite{Georgiou:2004wu,Georgiou:2004by}, Higgs boson 
\cite{Dixon:2004za}, massive gauge bosons \cite{Bern:2004ba}, photons \cite{Ozeren:2005mp}  and 
even massive fermions \cite{Badger:2005zh}.
In the case of pure gluonic processes, helicity amplitudes involving up to eight particles 
have been computed. For those involving also massless fermions 
calculations have been performed up to six particles \cite{Luo:2005rx,Luo:2005my}.
Furthermore, recent progress has been done to extend the validity of the recursion relations to 
one-loop amplitudes in QCD \cite{Bern:2005hs,Bern:2005ji,Bern:2005cq}.

In this paper, we present the first calculation of the full set of helicity amplitudes involving 
a quark-antiquark pair and five-gluons, needed for the computation of tree-level five-jet 
cross-sections, and an ingredient for the real part of NLO(NNLO) results for four(three)-jet 
observables in hadronic collisions.

This paper is organized as follows: in Section 2 we review the main ingredients of the BCFW 
formulation and recall its limitations when fermions are present. In Section 3 we introduce the 
main results for the complete set of $q\bar{q}5g$ helicity amplitudes, while in Section 4 we 
present our conclusions. 

\section{Color decomposition, helicity and BCFW}
\label{sec:review}

The color decomposition for a $q\bar{q}$ pair and $n$ gluons at tree level is particularly 
simple. The amplitude $M_n^{(0)}$ can be written in terms of the partial amplitude $A_n^{(0)}$ as 
\cite{Mangano:1990by}
\begin{equation}
\label{color}
M_n^{(0)}({k_i,\lambda_i,a_i})=g^{n-2}\sum_{\sigma\in S_{n-2}} (T^{a_{\sigma(3)}}\dotsc 
T^{a_{\sigma(n)}}){}_{\bar{j}_2}^{{i_1}}\, \, A_n^{(0)} (1_q^{\lambda_1},2_{\bar{q}}^{\lambda_2},
\sigma(3^{\lambda_3}),\dotsc,\sigma(n^{\lambda n})) \, ,
\end{equation}
where $ S_{n-2}$ is the group of permutations of $n-2$ symbols, with $1$ representing the 
quark with 
color $i_1$ and $2$ the antiquark with color $\bar{j}_2$. The upper-index $\lambda_l$ indicates 
the helicity of particle $l$ carrying momentum $k_l$.  The normalization for the color matrices 
in the fundamental representation is ${\rm Tr}\left( T^a T^b\right)=\delta^{ab}$.

In the framework of the helicity formalism \cite{Xu:1986xb,Kleiss:1985yh,Gunion:1985vc}, with the 
spinors denoted as 
\begin{equation}
\label{spinors}
\vert i^{\pm}\rangle=\vert k_i^{\pm}\rangle=\psi_{\pm} (k_i)
\qquad \qquad \langle i^{\pm}\vert=\langle k_i^ {\pm}\vert=\overline{\psi_{\pm} (k_i)} \, ,
\end{equation}
the partial amplitudes can be written in terms of the spinors inner-products
\begin{equation}
\label{inner}
\begin{split}
&\langle ij\rangle=\langle i^-\vert j^+\rangle=\overline{\psi_{-}}(k_i) \psi_{+} (k_j) \\
&[ij]=\langle i^+ \vert j^-\rangle=\overline{\psi_{+}} (k_i) \psi_{-} (k_j) \, ,
\end{split}
\end{equation}
and a few simple combinations of them, like
\begin{eqnarray}
\langle i \vert p_a\vert j] \equiv \langle i a \rangle [a j] \nn \\
\langle i \vert p_a p_b\vert j\rangle \equiv \langle i a \rangle [a b]\langle b j \rangle  \, .
\end{eqnarray}
In our convention all particles are considered to be outgoing and, following the QCD literature 
\cite{Mangano:1990by,Dixon:1996wi}, we fix the sign of the inner products such that 
$\langle ij \rangle [ji] = s_{ij}$\footnote{When comparing with results obtained using the 
string-like conventions just notice that $[ij]$ carries the opposite sign}.

The BCFW recurrence relation is based on the analytical properties of the amplitude when the spinors of two external legs  (denoted by $j$ and $l$) are shifted as 
\begin{equation}\label{bcfseg}
\begin{split}
&\vert \hat{j} \rangle = \vert j \rangle  \\
&\vert \hat{j}]= \vert j] - z \vert l] \\
&\vert \hat{l} \rangle = \vert l \rangle + z \vert j \rangle \\
&\vert \hat{l}]= \vert l ] \,.
\end{split}
\end{equation}

After this shift, the BCFW formula simply reads
\begin{equation}\label{BCF}
\begin{split}
A_n^{(0)} (1^{\lambda_1},\dotsc,n^{\lambda_n})= \sum_{r,s}\sum_{\lambda =\pm} 
A_{s-r+2}^{(0)} (r^{\lambda_r},\dotsc,\hat{j}^{\lambda_j},\dotsc,s,-\hat{K}_{r s}^{\lambda}) \\
\frac{1}{K^2_{r s}}\, A_{n-s+r}^{(0)} (\hat{K}_{r s}^{-\lambda},(s+1)^{\lambda_{(s+1)}},
\dotsc,\hat{l}^{\lambda_l},\dotsc,(r-1)^{\lambda_{(r-1)}}) \, , 
\end{split}
\end{equation}
where $K_{rs}= k_r +\dotsc+k_j \dotsc+k_s $ and the (complex) shift variable $z$ takes the 
value
\begin{equation}
\label{zrs}
z_{rs}=-\frac{(K_{rs})^2}{\langle j \vert K_{rs}\vert l]} \, .
\end{equation}

At this point we should make a  few remarks on Eq.(\ref{BCF}). First of all, each term is the 
product of two helicity amplitudes with a fewer number of particles and a propagator. The sum 
over $r$ and $s$ is not actually a sum over all of their 
possible values, but instead over all the possible configurations where the $j$ 
particle belongs to one of the amplitudes and the $l$ particle to the other one. For future 
reference, we shall call the amplitude with the $j$ particle ``upper amplitude'' and the one  
including the $l$ particle ``lower amplitude''. 
There are certain restrictions to the massless particles that can be used as the reference 
lines $j$ and $l$; 
in general they can not be chosen as $(\lambda_j,\lambda_l)=(+,-)$. 
Furthermore, quarks and antiquarks of the same flavor can not be chosen if they are adjacent 
and for adjacent quarks and gluons the helicities should better be opposite \cite{Luo:2005rx}.

We should also notice that the sum includes  amplitudes involving only 3 on-shell partons.
Because of helicity conservation these amplitudes would vanish 
if the momenta were not shifted. It is straightforward to show 
that, after the shift in Eq.(\ref{bcfseg}), only the 3 parton upper MHV 
and the 3-parton lower $\overline{\rm MHV}$ amplitudes become non-zero. 
Therefore, the $ggg$ and $q\bar{q}g$ MHV amplitudes \cite{Parke:1986gb}, 
which with our phase conventions read
\begin{eqnarray}
A _3^{(0)} (1^{+}_g,2^{-}_g,3^{-}_g)= \frac{\langle 2 3\rangle^3}
{\langle 1 2\rangle \langle 3 1\rangle} \nn \\
A _3^{(0)} (1^{+}_q,2^{-}_{\bar{q}},3^{-}_g)= \frac{\langle 2 3\rangle^2}{\langle 2 1\rangle }  \, ,
\end{eqnarray}
are the key ingredients of the recursion relations. 
The corresponding $\overline{\rm MHV}$ amplitudes can be obtained from those above by 
flipping the helicities 
applying
parity inversion ($\langle i j\rangle \rightarrow [ji]$ and an extra factor of -1 for each pair of 
quarks participating) and charge conjugation plus reflection and cyclic symmetries of the 
amplitudes.
Using the recursion relations it is possible to construct the tree level amplitude for $n$-partons 
just by conveniently iterating these building blocks.

\section{$q\bar{q}5g$ helicity amplitudes}
\label{sec:amplitudes}

There are in principle $2^7$ different helicity amplitudes for this process but half of them,
those with the quark and the antiquark carrying the same helicity, are trivially vanishing 
for massless particles. Furthermore, 
it is enough to present the results for one of the two possible combinations of
$q\bar{q}$ helicities (we choose here $1_q^+ \, 2_{\bar{q}}^-$); 
the other can be obtained by parity and charge conjugation. Out of the remaining 32 amplitudes, 
those with all gluons with the same helicity (two) are vanishing and other 10 are either MHV 
or $\overline{\rm MHV}$ amplitudes \cite{Parke:1986gb}, simply  reading
\begin{eqnarray}
A _7^{(0)} (1^{+}_q,2^{-}_{\bar{q}},3^+_g,\dotsc,i^-_g\dotsc,7^+_g)= 
\frac{\langle 2i\rangle^3 \langle 1i\rangle} 
{\prod^7_{l=1} \langle l\medspace \medspace l+1\rangle} \nn \, \\
A _7^{(0)} (1^{+}_q,2^{-}_{\bar{q}},3^-_g,\dotsc,i^+_g\dotsc,7^-_g)= -\frac{[1i]^3  [2i]}
{\prod^7_{l=1} [ l\medspace \medspace l+1]}\, ,
\end{eqnarray}
where $i$ represents the gluon with the opposite helicity to the others.

Therefore, there are only 10 non-trivial NMHV, corresponding to three gluons with helicity plus
and two with helicity minus, amplitudes to be computed. Again, the 10 $\overline{\rm NMHV}$ 
amplitudes 
can be obtained by the discrete symmetries P and C. 
Further simplifications in the number of independent amplitudes could be achieved by applying 
supersymmetric 
relations. We rather present the explicit results for those 10 amplitudes in order
to provide the most compact expressions for direct use.

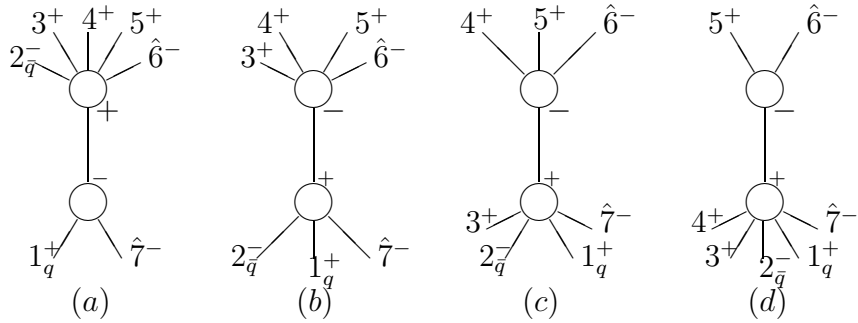
\begin{figure}[!htp]
\setlength{\unitlength}{0.10mm}
\begin{picture}(300,400)(-280,-70)
\put(150,75){\circle{50}}
\put(150,225){\circle{50}}
\put(150,102){\line(0,1){97}}
\put(160,185){$+$}
\put(155,105){$_-$}
\put(450,75){\circle{50}}
\put(450,225){\circle{50}}
\put(450,102){\line(0,1){97}}
\put(460,185){$-$}
\put(455,105){$_+$}
\put(750,75){\circle{50}}
\put(750,225){\circle{50}}
\put(750,102){\line(0,1){97}}
\put(760,185){$-$}
\put(755,105){$_+$}
\put(1050,75){\circle{50}}
\put(1050,225){\circle{50}}
\put(1050,102){\line(0,1){97}}
\put(1060,185){$-$}
\put(1055,105){$_+$}
\put(105,0){\line(3,5){31}}
\put(195,0){\line(-3,5){31}}
\put(70,-10){$1_q^+$}
\put(205,-10){$\hat{7}^-$}
\put(1005,300){\line(3,-5){31}}
\put(1095,300){\line(-3,-5){31}}
\put(975,300){$5^+$}
\put(1105,300){$\hat{6}^-$}
\put(375,0){\line(1,1){55}}
\put(450,0){\line(0,1){49}}
\put(525,0){\line(-1,1){55}}
\put(340,-10){$2_{\bar{q}}^-$}
\put(442,-25){$1_q^+$}
\put (535,-10){$\hat{7}^-$}
\put(675,300){\line(1,-1){55}}
\put(750,300){\line(0,-1){48}}
\put(825,300){\line(-1,-1){55}}
\put(645,300){$4^+$}
\put(742,305){$5^+$}
\put(835,300){$\hat{6}^-$}
\put(705,0){\line(3,5){31}}
\put(795,0){\line(-3,5){31}}
\put(680,40){\line(2,1){45}}
\put(820,40){\line(-2,1){45}}
\put(670,-10){$2_{\bar{q}}^-$}
\put(805,-10){$1_q^+$}
\put(652,35){$3^+$}
\put(830,35){$\hat{7}^-$}
\put(405,300){\line(3,-5){31}}
\put(495,300){\line(-3,-5){31}}
\put(380,260){\line(2,-1){45}}
\put(520,260){\line(-2,-1){45}}
\put(375,300){$4^+$}
\put(505,300){$5^+$}
\put(353,255){$3^+$}
\put(530,255){$\hat{6}^-$}
\put(1005,0){\line(3,5){31}}
\put(1095,0){\line(-3,5){31}}
\put(980,40){\line(2,1){45}}
\put(1120,40){\line(-2,1){45}}
\put(1047,0){\line(0,1){49}}
\put(970,-10){$3^+$}
\put(1105,-10){$1_q^+$}
\put(952,35){$4^+$}
\put(1130,35){$\hat{7}^-$}
\put(1042,-25){$2_{\bar{q}}^-$}
\put(105,300){\line(3,-5){31}}
\put(195,300){\line(-3,-5){31}}
\put(80,260){\line(2,-1){45}}
\put(220,260){\line(-2,-1){45}}
\put(150,300){\line(0,-1){48}}
\put(75,300){$3^+$}
\put(205,300){$5^+$}
\put(45,255){$2_{\bar{q}}^-$}
\put(230,255){$\hat{6}^-$}
\put(142,305){$4^+$}
\put(127,-70){$(a)$}
\put(427,-70){$(b)$}
\put(727,-70){$(c)$}
\put(1027,-70){$(d)$}
\end{picture}
\caption{Diagrams contributing to $A_7^{(0)}(1_q^+,2_{\bar{q}}^-,3^+,4^+,5^+,6^-,7^-)$} 
\label{7p}
\end{figure}

The use of the BCFW formula for $q\bar{q}5g$ amplitudes involves the appearance of, at most, 
four different arrangements in the recursion. This is shown in Fig.(\ref{7p}) for 
the $(1_q^+,2_{\bar{q}}^-,3^+,4^+,5^+,6^-,7^-)$ helicity configuration, where gluons 6 and 7
were chosen as lines $j$ and $l$ in Eq.(\ref{bcfseg}), respectively. As known, a smart election
for the reference lines can result into more compact expressions for the amplitudes. 

In this case  
diagram (a), denoted as $(2,3,4,5,\hat{6} | \hat{7},1)$ vanishes because the two fermions 
appear in different subamplitudes, fixing the helicity of the propagator
and selecting a 3-parton lower MHV amplitude. The remaining diagrams ( $(3,4,5,\hat{6} | \hat{7},1,2)$, 
$(4,5,\hat{6} | \hat{7},1,2,3)$, and  $(5,\hat{6} | \hat{7},1,2,3,4)$) are simply products of 
MHV and/or $\overline{\rm MHV}$ amplitudes, so each of them contributes with a single term. 
Our final result reads
\begin{equation}
\label{amp1}
\begin{split}
&A_7^{(0)}(1_q^+,2_{\bar{q}}^-,3^+,4^+,5^+,6^-,7^-)=-\frac{\langle6\vert7+2\vert1]^3}
{s_{712}[71][12]\langle34\rangle \langle 45\rangle
 \langle 56\rangle \langle 3\vert1+2\vert7]} \\
&+\frac{\langle2\vert6+7\vert5]^3}{s_{567}\langle12\rangle \langle23\rangle \langle34\rangle
 [56][67] \langle4\vert5+6\vert7] }
-\frac{\langle2\vert(1+3)(4+5)\vert6\rangle^3}{s_{123}s_{456} \langle12\rangle \langle23\rangle  
\langle45\rangle \langle56\rangle \langle3\vert1+2\vert7]\langle4\vert5+6\vert7]} 
\, ,
\end{split}
\end{equation}
where $s_{ijk}=(p_i+p_j+p_k)^2$. Factorization properties of the amplitudes in the collinear 
and soft limits provide stringent consistency checks to them. For example, when gluon 
6 is soft, Eq.(\ref{amp1}) becomes
\begin{equation} 
A_7^{(0)}(1_q^+,2_{\bar{q}}^-,3^+,4^+,5^+,6^-,7^-) \xrightarrow{k_6 \to 0} 
\left( \frac{- \langle 27 \rangle^3}{\langle 12 \rangle \langle 23 \rangle \langle 34 \rangle 
\langle 45 \rangle \langle 57 \rangle} \right) (-\frac{[51]}{[56][61]}) \, ,
\end{equation}
i.e., just the product of the six particle amplitude 
$A_6^{(0)}(1_q^+,2_{\bar{q}}^-,3^+,4^+,5^+,7^-)$ 
times the eikonal factor for the emission of a soft gluon with negative helicity.

In our search for compact expressions, we can show that the NMHV 7-parton amplitudes have at 
most six terms. In any amplitude, diagrams like (a) contribute with (at most) three terms 
(since it involves a 6 parton NMHV amplitude \cite{Mangano:1990by,Luo:2005rx}) 
and each of the other diagrams adds a single term, if they don't vanish. 

The amplitude corresponding to the ordering $++-+-$ on the gluon helicities is computed 
choosing $j=3$ and $l=4$ as reference lines, resulting 
\begin{equation}
\begin{split}
&A_7^{(0)}(1_q^+,2_{\bar{q}}^-,3^+,4^+,5^-,6^+,7^-)= 
\\&
-\frac{\langle5\vert3+4\vert1]^3\langle57\rangle^4}{[12]\langle34\rangle\langle45\rangle\langle56
\rangle\langle67\rangle\langle5\vert6+7\vert1]\langle3\vert(1+2)(6+7)\vert5\rangle\langle7
\vert(1+2)(3+4)\vert5\rangle}
\\&
+\frac{\langle27\rangle^3\langle5\vert3+4\vert6]^4}{s_{712}s_{345}\langle12\rangle
\langle34\rangle\langle45\rangle\langle3\vert4+5\vert6]\langle2\vert7+1\vert6]\langle7
\vert(1+2)(3+4)\vert5\rangle}
\\&
-\frac{\langle25\rangle^3[16]^3\langle5\vert7+1\vert6]}{s_{671}\langle23\rangle\langle34\rangle
\langle45\rangle[67][71]\langle2\vert7+1\vert6]\langle5\vert6+7\vert1]}
\\&
-\frac{\langle27\rangle^3[46]^4}{s_{456}\langle12\rangle\langle23\rangle[45][56]\langle3\vert
4+5\vert6]\langle7\vert6+5\vert4]}
\\&
+\frac{\langle2\vert1+3\vert4]^3\langle57\rangle^4}{s_{123}s_{567}\langle12\rangle
\langle23\rangle\langle56\rangle\langle67\rangle\langle7\vert6+5\vert4]\langle3\vert(1+2)(6+7)
\vert5\rangle} \, ,
\end{split}
\end{equation}
where we have splitted the result in the following order: the first three terms come from 
$(6,7,1,2,\hat{3} | \hat{4},5)$, the next single term from $(7,1,2,\hat{3} | \hat{4},5,6)$ 
and the last 
term from $(1,2,\hat{3} | \hat{4},5,6,7)$. The contribution from $(2,\hat{3} | \hat{4},5,6,7,1)$ 
 vanishes because the ``fermion propagator'' selects an upper $\overline{\rm MHV}$ amplitude.

The amplitude corresponding to the ordering $++--+$ on the gluon helicities is also computed 
choosing $j=3$ and $l=4$ as reference lines, obtaining
\begin{equation}
\begin{split}
&A_7^{(0)}(1_q^+,2_{\bar{q}}^-,3^+,4^+,5^-,6^-,7^+)=
\\&
+\frac{[17]^2\langle25\rangle^3\langle5\vert1+6\vert7]}{s_{671}\langle23\rangle\langle34\rangle
\langle45\rangle[67]\langle5\vert6+7\vert1]\langle2\vert7+1\vert6]}
\\&
-\frac{\langle2\vert(7+1)(3+4)\vert5\rangle^3\langle1\vert(2+7)(3+4)\vert5\rangle}
{s_{712}s_{345}\langle71\rangle\langle12\rangle\langle34\rangle\langle45\rangle
\langle3\vert4+5\vert6]\langle2\vert7+1\vert6]\langle7\vert(1+2)(3+4)\vert5\rangle}
\\&
-\frac{\langle5\vert4+3\vert1]^3\langle56\rangle^3}{[12]\langle34\rangle\langle45\rangle
\langle67\rangle\langle5\vert6+7\vert1]\langle3\vert(1+2)(6+7)\vert5\rangle\langle7\vert(1+2)
(3+4)\vert5\rangle}
\\&
+\frac{\langle2\vert5+6\vert4]^3\langle1\vert5+6\vert4]}{s_{456}\langle71\rangle\langle12\rangle
\langle23\rangle[45][56]\langle3\vert4+5\vert6]\langle7\vert5+6\vert4]}
\\&
+\frac{\langle2\vert1+3\vert4]^3\langle56\rangle^3}{s_{567}s_{123}\langle12\rangle\langle23
\rangle\langle67\rangle\langle7\vert5+6\vert4]\langle3\vert(1+2)(6+7)\vert5\rangle} \, .
\end{split}
\end{equation}
The first three terms come from 
$(6,7,1,2,\hat{3} | \hat{4},5)$, the fourth from $(7,1,2,\hat{3} | \hat{4},5,6)$ and the fifth
 from $(1,2,\hat{3} | \hat{4},5,6,7)$. Again, because of the ``fermion propagator'' 
 and the helicities of the particles involved, $(2,\hat{3} | \hat{4},5,6,7,1)$ has a 
 null contribution to this result.

\newpage 

The amplitude corresponding to the ordering $+-++-$  is obtained selecting $j=5$ and $l=6$ as 
reference lines, 
\begin{equation}
\begin{split}
&A_7^{(0)}(1_q^+,2_{\bar{q}}^-,3^+,4^-,5^+,6^+,7^-)=
\\&
-\frac{\langle24\rangle^3\langle7\vert5+6\vert1]^3\langle4\vert(2+3)(5+6)\vert7\rangle}
{s_{234}s_{567}\langle23\rangle\langle34\rangle\langle56\rangle\langle67\rangle
\langle4\vert2+3\vert1] \langle5\vert6+7\vert1]\langle2\vert(3+4)(5+6)\vert7\rangle}
\\&
-\frac{\langle27\rangle^3\langle7\vert5+6\vert3]^4}{\langle12\rangle[34]\langle56\rangle
\langle67\rangle\langle7\vert1+2\vert3]\langle7\vert5+6\vert4]\langle5\vert(3+4)(1+2)\vert7\rangle
\langle2\vert(3+4)(5+6)\vert7\rangle}
\\&
-\frac{[13]^3\langle47\rangle^4}{s_{123}[12]\langle45\rangle\langle56\rangle\langle67\rangle
\langle4\vert2+3\vert1]\langle7\vert1+2\vert3]}
\\&
-\frac{\langle24\rangle^3[16]^3\langle4\vert7+1\vert6]}{s_{671}\langle23\rangle
\langle34\rangle\langle45\rangle[67][71]\langle2\vert7+1\vert6]\langle5\vert6+7\vert1]}
\\&
-\frac{\langle27\rangle^3\langle4\vert5+3\vert6]^4}{s_{712}s_{345}\langle12\rangle\langle34
\rangle\langle45\rangle\langle3\vert4+5\vert6]\langle2\vert7+1\vert6]\langle5\vert(3+4)(1+2)
\vert7\rangle}
\\&
-\frac{\langle27\rangle^3[56]^3}{s_{456}\langle12\rangle\langle23\rangle[45]
\langle3\vert4+5\vert6]\langle7\vert6+5\vert4]} \, .
\end{split}
\end{equation}
This formula was written in the following order: the first three terms come from 
$(1,2,3,4,\hat{5} | \hat{6},7)$, the next one from $(2,3,4,\hat{5} | \hat{6},7,1)$, a single
term from $(3,4,\hat{5} | \hat{6},7,1,2)$ and the last one comes from $(4,\hat{5} | \hat{6},7,1,2,3)$.

The ordering $+-+-+$ can be obtaining setting $j=6$ and $l=7$, 
\begin{equation}
\begin{split}
&A_7^{(0)}(1_q^+,2_{\bar{q}}^-,3^+,4^-,5^+,6^-,7^+)=
\\&
\frac{\langle 16\rangle \langle 24\rangle^3 \langle 4\vert 2+3\vert5] 
\langle 6\vert 7+1\vert 5]^3}{s_{671}s_{234} \langle 23 \rangle \langle 34 \rangle 
\langle 67 \rangle \langle 71 \rangle \langle 1 \vert 6+7 \vert 5] \langle 2 \vert 3+4 \vert 5] 
\langle 6 \vert (7+1) (2+3) \vert 4 \rangle} 
\\&
+\frac{\langle 16 \rangle \langle 26 \rangle^3 [35]^4}{s_{345} \langle 67 \rangle 
\langle 71 \rangle \langle 12 \rangle [34] [45] \langle 6 \vert 4+5 \vert3] 
\langle 2 \vert 3+4 \vert 5]}
\\&
+\frac{\langle 16 \rangle \langle 46 \rangle^4 \langle 6 \vert 7+1 \vert 3]^3}{\langle 45 \rangle 
\langle 56 \rangle \langle 67 \rangle \langle 71 \rangle \langle 6 \vert 4+5 \vert 3]  
\langle 6 \vert 7+1 \vert 2] 
\langle 6 \vert (4+5) (2+3) \vert 1 \rangle \langle 6 \vert (7+1) (2+3) \vert 4 \rangle}
\\&
+\frac{[27][17]^2 \langle 46 \rangle^4}{s_{712} [12] \langle 34 \rangle \langle 45 \rangle 
\langle 56 \rangle \langle 3 \vert 1+2 \vert7] \langle 6 \vert 7+1 \vert 2]}
\\&
+\frac{\langle 1 \vert 2+3 \vert7] \langle 2 \vert 1+3 \vert7]^3 \langle 46 \rangle^4}{s_{123} 
s_{456} \langle 12 \rangle \langle 23 \rangle \langle 45 \rangle \langle 56 \rangle 
\langle 4 \vert 5+6 \vert 7] \langle 3 \vert 1+2 \vert 7] 
\langle 6 \vert (4+5) (2+3) \vert 1 \rangle}
\\&
+\frac{\langle 14 \rangle \langle 24 \rangle^3 [57]^4}{s_{567} \langle 12 \rangle 
\langle 23 \rangle \langle 34 \rangle [56][67] \langle 4\vert 5+6 \vert7] 
\langle 1 \vert 6+7 \vert 5]} \, ,
\end{split}
\end{equation}
where the first three terms come from 
$(2,3,4,5,\hat{6} | \hat{7},1)$, the fourth term from $(3,4,5,\hat{6} | \hat{7},1,2)$, the fifth 
 from $(4,5,\hat{6} | \hat{7},1,2,3)$ and the sixth term from $(5,\hat{6} | \hat{7},1,2,3,4)$.

\newpage 

The amplitude corresponding to the ordering $+--++$  is obtained by choosing $j=6$ and $l=7$ as 
reference lines, 
\begin{equation}
\begin{split}
&A_7^{(0)}(1_q^+,2_{\bar{q}}^-,3^+,4^-,5^-,6^+,7^+)=
\\&
+\frac{\langle 4 \vert (2+3) (6+7) \vert 1 \rangle (s_{671})^2 \langle 24 \rangle^3}
{s_{234} \langle 23 \rangle \langle 34 \rangle \langle 67 \rangle \langle 71 \rangle 
\langle 1 \vert 6+7 \vert 5] \langle 2 \vert 3+4 \vert 5] 
\langle 4 \vert (2+3) (7+1) \vert 6\rangle} 
\\&
+\frac{\langle 1 \vert 4+5 \vert 3]  \langle 2 \vert 4+5 \vert 3]^3}{s_{345} \langle 67 \rangle
\langle 71 \rangle \langle 12 \rangle [34] [45] \langle 2\vert 3+4 \vert 5] 
\langle 6 \vert 5+4 \vert 3]} 
\\&
-\frac{\langle 16 \rangle \langle 45 \rangle^3 \langle 6 \vert 7+1 \vert3]^3}
{\langle 56 \rangle 
\langle 67 \rangle \langle 71 \rangle \langle 6 \vert 7+1 \vert 2] \langle 6 \vert 5+4 \vert 3] 
\langle 6 \vert (4+5) (2+3) \vert 1 \rangle \langle 4 \vert (2+3) (7+1) \vert 6 \rangle}
\\&
+\frac{[27][17]^2 \langle 45 \rangle^3}{s_{712} [12] \langle 34 \rangle \langle 56 \rangle
\langle 3 \vert 1+2 \vert 7] \langle 6 \vert 7+1 \vert 2]}
\\&
+\frac{\langle 1 \vert 2+3 \vert 7] \langle 2 \vert 1+3 \vert 7]^3 \langle 45 \rangle^3}
{s_{123} s_{456} \langle 12 \rangle \langle 23 \rangle \langle 56 \rangle 
\langle 3 \vert 1+2 \vert 7] \langle 4\vert 5+6 \vert 7]
\langle 6 \vert (4+5) (2+3) \vert 1\rangle}
\\&
+\frac{\langle 14 \rangle \langle 24 \rangle^3 [67]^3}{s_{567} \langle 12 \rangle 
\langle 23 \rangle \langle 34 \rangle [56] \langle 4 \vert 5+6 \vert 7] 
\langle 1 \vert 6+7 \vert 5]} \, ,
\end{split}
\end{equation}
where we have written the result in the following order: the first three terms come from 
$(2,3,4,5,\hat{6} | \hat{7},1)$, a single term from $(3,4,5,\hat{6} | \hat{7},1,2)$, the next 
term from $(4,5,\hat{6} | \hat{7},1,2,3)$ and the last term from $(5,\hat{6} | \hat{7},1,2,3,4)$.

The ordering $-+++-$, using $j=7$ and $l=1$, results
\begin{equation}
\begin{split}
&A_7^{(0)}(1_q^+,2_{\bar{q}}^-,3^-,4^+,5^+,6^+,7^-)=
\\&
\frac{\langle 7 \vert 5+6 \vert 4]^3 \langle 27 \rangle^3}{\langle 12 \rangle [34] 
\langle 56 \rangle \langle 67 \rangle \langle 7 \vert 1+2 \vert 3] 
\langle 2 \vert (3+4) (5+6) \vert 7 \rangle \langle 7 \vert (1+2) (3+4) \vert 5 \rangle } 
\\&
+\frac{\langle 3 \vert 4+5 \vert 6]^3 \langle 27 \rangle^3}{s_{712} s_{345} 
\langle 12 \rangle \langle 34 \rangle \langle 45 \rangle \langle 2 \vert 7+1 \vert 6] 
\langle 7 \vert (1+2) (3+4) \vert 5 \rangle} 
\\&
-\frac{\langle 7 \vert 1+3 \vert 2] \langle 7 \vert 2+3 \vert 1]^3}{s_{123} [12] [23] 
\langle 45 \rangle \langle 56 \rangle \langle 67 \rangle \langle 7 \vert 1+2 \vert 3] 
\langle 4 \vert 3+2 \vert 1] } 
\\&
-\frac{\langle 7 \vert (5+6) (2+4) \vert 3 \rangle \langle 7 \vert 5+6 \vert 1]^3 
\langle 23 \rangle^3}{s_{234}s_{567} \langle 23 \rangle \langle 34 \rangle 
\langle 56 \rangle \langle 67 \rangle \langle 4 \vert 3+2 \vert 1] \langle 5 \vert 6+7 \vert 1] 
\langle 7 \vert (5+6) (3+4) \vert 2 \rangle} 
\\&
-\frac{ \langle 3 \vert 7+1 \vert 6] \langle 23 \rangle^2 [16]^3}{s_{671} \langle 34 \rangle 
\langle 45 \rangle [67][71] \langle 2 \vert 7+1 \vert 6] \langle 5 \vert 6+7 \vert 1]} \, .
\end{split}
\end{equation}
The different contributions appear in the following order: the first two terms come from 
$(3,4,5,6,\hat{7} | \hat{1},2)$, the third term comes from $(4,5,6,\hat{7} | \hat{1},2,3)$, 
the fourth term from $(5,6,\hat{7} | \hat{1},2,3,4)$ and the last term from $(6,\hat{7} | \hat{1},2,3,4,5)$.

\newpage 

The amplitude corresponding to the ordering $-+-++$  is obtained by selecting $j=5$ and $l=6$ as 
reference lines, 
\begin{equation}
\begin{split}
&A_7^{(0)}(1_q^+,2_{\bar{q}}^-,3^-,4^+,5^-,6^+,7^+)=
\\&
-\frac{\langle 15 \rangle [24] \langle 5 \vert 2+3 \vert 4]^3}
{s_{234} [23] [34] \langle 56 \rangle \langle 67 \rangle \langle 71 \rangle 
\langle 5 \vert 3+4 \vert 2] \langle 1 \vert 2+3\vert 4]} 
\\&
+ \frac{\langle 5 \vert 6+7 \vert 2] \langle 5 \vert 6+7 \vert 1]^2 \langle 35 \rangle^4}
{[12] \langle 34 \rangle \langle 45 \rangle \langle 56 \rangle \langle 67 \rangle 
\langle 5 \vert 3+4 \vert 2] \langle 3 \vert (1+2) (6+7) \vert 5 \rangle  
\langle 5 \vert (3+4) (1+2) \vert 7 \rangle }  
\\&
+\frac{\langle 13 \rangle \langle 23 \rangle^2 \langle 5 \vert 6+7 \vert 4]^4}{s_{567}s_{123} 
\langle 12 \rangle \langle 56 \rangle \langle 67 \rangle \langle 7 \vert 6+5 \vert 4] 
\langle 1 \vert 2+3 \vert 4] \langle 3 \vert (1+2) (6+7) \vert 5 \rangle } 
\\&
+\frac{\langle 1 \vert 7+2 \vert6] \langle 2 \vert 7+1 \vert 6]^2 \langle 35 \rangle^4}
{s_{712} s_{345} \langle 71 \rangle \langle 12 \rangle \langle 34 \rangle \langle 45 \rangle 
\langle 3 \vert 4+5 \vert 6] \langle 5 \vert (3+4) (1+2) \vert 7 \rangle} 
\\&
+\frac{\langle 13 \rangle \langle 23 \rangle^2 [46]^4}{s_{456} \langle 71 \rangle 
\langle 12 \rangle [45] [56] \langle 3 \vert 4+5\vert 6] \langle 7 \vert 5+6 \vert 4]} \, ,
\end{split}
\end{equation}
where we have written the result in the following order: the three terms from 
$(1,2,3,4,\hat{5} | \hat{6},7)$, a single 
term from $(3,4,\hat{5} | \hat{6},7,1,2)$ and the single term from 
$(4,\hat{5} | \hat{6},7,1,2,3)$. The contribution from 
$(2,3,4,\hat{5} | \hat{6},7,1)$ vanishes because of 
the ``fermion propagator'' and the helicities of the particles involved.

The ordering $-++-+$  is solved choosing $j=3$ and $l=4$ 
\begin{equation}
\begin{split}
&A_7^{(0)}(1_q^+,2_{\bar{q}}^-,3^-,4^+,5^+,6^-,7^+)= 
\\&
\frac{\langle 16 \rangle \langle 3 \vert 4+5 \vert2] \langle 6 \vert (7+1) (4+5) \vert 3 
\rangle^3}{s_{345}s_{671} \langle 34 \rangle \langle 45 \rangle \langle 67 \rangle \langle 71 
\rangle \langle 5 \vert 4+3 \vert 2] \langle 6 \vert 7+1 \vert 2] \langle 1 \vert (6+7) (4+5) 
\vert 3 \rangle} 
\\&
+\frac{[27][17]^2\langle 36 \rangle^4}{s_{712}[12] \langle 34 \rangle \langle 45 \rangle 
\langle 56 \rangle \langle 6 \vert 7+1 \vert 2] \langle 3 \vert 1+2 \vert 7]} 
\\&
- \frac{\langle 13 \rangle \langle 23 \rangle^2 \langle 3 \vert 4+5 \vert 7]^4}
{\langle 12 \rangle \langle 34 \rangle \langle 45 \rangle [67] \langle 3 \vert 4+5 \vert 6] 
\langle 3 \vert 1+2 \vert 7] \langle 3 \vert (1+2) (6+7) \vert 5 \rangle \langle 1 \vert (6+7) 
(4+5) \vert 3 \rangle} 
\\&
+\frac{\langle 13 \rangle \langle 23 \rangle^2 [45]^3}{s_{456} \langle 71 \rangle 
\langle 12 \rangle [56] \langle 3 \vert 4+5 \vert 6] \langle 7 \vert 5+6 \vert 4]} 
\\&
+\frac{\langle 13 \rangle \langle 23 \rangle^2 \langle 6 \vert 5+7 \vert4]^4}{s_{567}
s_{123} \langle 12 \rangle \langle 56 \rangle \langle 67 \rangle \langle 1 \vert 2+3 \vert 4] 
\langle 7 \vert 5+6 \vert 4] \langle 3 \vert (1+2) (6+7) \vert 5 \rangle} 
\\&
-\frac{\langle 16 \rangle [24] \langle 6 \vert 2+3 \vert 4]^3}{s_{234}[23][34] 
\langle 56 \rangle \langle 67 \rangle \langle 71 \rangle \langle 1 \vert 2+3 \vert4] 
\langle 5 \vert 4+3\vert2]} \, .
\end{split}
\end{equation}
This expression was splitted in the following order: the first three terms come from 
$(6,7,1,2,\hat{3} | \hat{4},5)$, the fourth from $(7,1,2,\hat{3} | \hat{4},5,6)$, the fifth 
term from $(1,2,\hat{3} | \hat{4},5,6,7)$ and the sixth term from $(2,\hat{3} | \hat{4},5,6,7,1)$.

Finally, the non-alternating amplitude $--+++$ (from $j= 4$ and $l=5$) turns out to be
\begin{equation}
\begin{split}
&A_7^{(0)}(1_q^+,2_{\bar{q}}^-,3^-,4^-,5^+,6^+,7^+)= 
\frac{\langle 1 \vert (2+7) (5+6) \vert4 \rangle \langle 2 \vert (7+1) (5+6) \vert 4 \rangle^2}
{s_{456}s_{712}\langle 71 \rangle \langle 12 \rangle \langle 45 \rangle \langle 56 \rangle 
\langle 6 \vert 5+4\vert 3] \langle 7 \vert 1+2 \vert 3]} 
\\&
- \frac{\langle 4\vert 1+3\vert2] \langle 4 \vert 2+3 \vert 1]^2}{s_{123}[12][23] 
\langle 45 \rangle \langle 56 \rangle \langle 67 \rangle \langle 7 \vert 1+2 \vert 3]} 
+ \frac{\langle 1 \vert 3+4 \vert5] \langle 2 \vert 3+4 \vert 5]^2}{s_{345} \langle 67 \rangle 
\langle 71 \rangle \langle 12 \rangle [34][45] \langle 6 \vert 5+4 \vert 3]} \, ,
\end{split}
\end{equation}
where we have written the result in the following order: the first two terms come from 
$(7,1,2,3,\hat{4} | \hat{5},6)$, and the last one from $(3,\hat{4} | \hat{5},6,7,1,2)$.
Contributions from $(1,2,3,\hat{4} | \hat{5},6,7)$ and $(2,3,\hat{4} | \hat{5},6,7,1)$ vanish
because of helicity conservation. 

As a byproduct of this calculation, several QED amplitudes can be obtained form our results, 
by turning any number of gluons into photons. In order to disguise gluons as photons one has 
to replace in Eq.(\ref{color}) the $SU(3)$ color group of 
QCD by the $U(1)$ group of QED.
The color decomposition for two quarks, $m$ photons and $r$ gluons \cite{DelDuca:1999pa} becomes
\begin{equation}
\begin{split}
M_n^{(0)}({k_i,\lambda_i,a_i})=g^{r} (\sqrt{2}eQ_{q})^m \sum_{\sigma\in S_{r}} 
(T^{a_{\sigma(3)}}\dotsc 
T^{a_{\sigma(r+2)}}){}_{\bar{j}_2}^{{i_1}}\, \, \\
A_n^{(0)} (1_q^{\lambda_1},2_{\bar{q}}^{\lambda_2},
\sigma_{g}(3^{\lambda_3}),\dotsc,\sigma_{g}((r+2)^{\lambda_{ (r+2)}}), \gamma_{r+3},\dotsc,\gamma_{n}) \, ,
\end{split}
\end{equation}
where $S_r$ is the group of permutations of $r$ symbols with $1$ being the quark with color $i_1$, 
$2$ the antiquark with color $\bar{j}_2$, and $\lambda_l$ represents the helicity of the $l$ 
particle carrying momentum $k_l$. This formula can be easily obtained from Eq.(\ref{color}) by 
converting the last $m$ gluons into photons. In order to make this conversion, one has to replace 
the factor $g T^a$ with $\sqrt{2}eQ_{q} I$, where $I$ is the identity matrix. Since 
$I$ commutes with the $T^a$ matrices, all possible photon-quark couplings contribute to the 
same color structure. Therefore to obtain the amplitudes involving photons, one has to sum over 
all the permutations where the photon ``moves'' in the amplitude while the gluons remain fixed.

From the previous results one can change any number of gluons into photons, using the procedure 
described in the previous paragraph, allowing to obtain the following QED/QCD amplitudes: 
$q \bar{q} 4g \gamma$,  $q \bar{q} 3g 2\gamma$, $q \bar{q} 2g 3\gamma$,  $q \bar{q} g 4\gamma$ and
 $q \bar{q}  5\gamma$ .

\section{Conclusions}
\label{sec:conclu}

In this paper we presented all seven parton tree level NMHV amplitudes involving a
fermionic pair and five gluons, obtained by use of  the BCFW recursion relations. 
 With the knowledge of these amplitudes  the full set of 
helicity amplitudes for 
the two quarks plus five gluons process is available. The trivial MHV amplitudes 
are giving by the Parke-Taylor formulae 
 and the NNMHV amplitudes are also $\overline{\rm NMHV}$ so 
they can be obtained by performing the adequate combination of P and C discrete symmetries
over our results. We should emphasize that the results presented in this paper have been 
checked in all possible collinear and soft limits,  setting and stringent test for the 
correctness of the amplitudes
\footnote{Stricktly speaking they are correct up to terms that must vanish 
in all possible soft and collinear limits, which are very unlikely to exist.}.

Furthermore, by making simple 
replacements in the color decomposition formula, one can obtain several seven 
parton 
QED/QCD amplitudes involving two quarks, $m$ photons and $r$ gluons (where $m+r=5$).

These amplitudes are a main ingredient for the calculation of multijets  cross sections
in hadronic colliders.
 As expected, 
we have  obtained very compact expressions for the amplitudes, allowing for a more convenient 
implementation 
in computer codes than those obtained from automatic tree level computations methods. 

\noindent {\bf  Acknowledgements.} 
We would like to thank Alejandro Daleo and Zoltan Kunszt for discussions.
This work has been partially supported by ANPCYT, UBACyT and CONICET.


\end{document}